\documentclass[usenatbib]{mn2e}

\usepackage{graphicx}
\usepackage{amssymb}
\usepackage{epsfig}

\title[Pdot]{Exploring accretion theory with X-ray binaries in the SMC}

\author[M.J. Coe et al.]{M. J.~Coe$^{1}$,
 V.A. ~McBride$^{1}$ \& R.H.D. ~Corbet$^{2}$ \\
$^{1}$ School of Physics and Astronomy, University of Southampton, SO17
1BJ, UK\\
$^{2}$ University of Maryland, Baltimore County, Mail Code 662, NASA Goddard Space Flight Center, Greenbelt, MD 20771, USA \\
}

\begin{document}

\date{21 August 2009}

\pagerange{\pageref{firstpage}--\pageref{lastpage}} \pubyear{2002}

\maketitle

\label{firstpage}

\begin{abstract}

The understanding of the accretion process on to compact objects in binary systems is an important part of modern astrophysics. Theoretical work, primarily that of Ghosh \& Lamb (1979), has made clear predictions for the behaviour of such systems which have been generally supported by observational results of considerably varying quality from galactic accreting pulsar systems. In this work a much larger homogeneous population of such objects in the Small Magellanic Cloud (SMC) is used to provide more demanding tests of the accretion theory. The results are extremely supportive of the theoretical models and provide useful statistical insights into the manner in which accreting pulsars behave and evolve.

\end{abstract}


\section{Introduction and background}

The Be/X-ray systems represent the largest sub-class of all High Mass X-ray Binaries (HMXB).  A survey of the literature reveals that of the $\sim$240 HMXBs known in our Galaxy and the Magellanic Clouds (Liu et al., 2005, 2006), $\ge$50\%
fall within this class of binary.  In fact, in recent years it has emerged that there is a substantial population of HMXBs in the SMC comparable in number to the Galactic population. Though unlike the Galactic population, all except one of the SMC HMXBs are Be star systems.  In these systems the orbit of the Be star
and the compact object, presumably a neutron star, is generally wide
and eccentric.  X-ray outbursts are normally associated with the
passage of the neutron star close to the circumstellar disk (Okazaki
\& Negueruela 2001), and generally are classified as Types I or II (Stella, White \& Rosner, 1986). The Type I outbursts occur periodically at the time of the periastron passage of the neutron star, whereas Type II outbursts are much more extensive and occur when the circumstellar material expands to fill most, or all of the orbit. This paper concerns itself with Type I outbursts. General reviews of such HMXB systems may
be found in Negueruela (1998), Corbet et al. (2008) and Coe et al. (2000, 2008).

This paper reports data acquired over 10 years using the Rossi X-ray Timing Explorer (RXTE) on the HMXB population of the SMC. During the period of these observations there have been many opportunities to study spin changes arising from accretion torques. This extremely homogeneous population permits the first high quality tests to be carried out of the work of Ghosh \& Lamb (1979) and Joss \& Rappaport (1984). The simplified source naming convention used in this work follows that established by Coe et al (2005).

\section{Observations}

\begin{figure}
\begin{center}
\includegraphics[width=70mm,angle=-0]{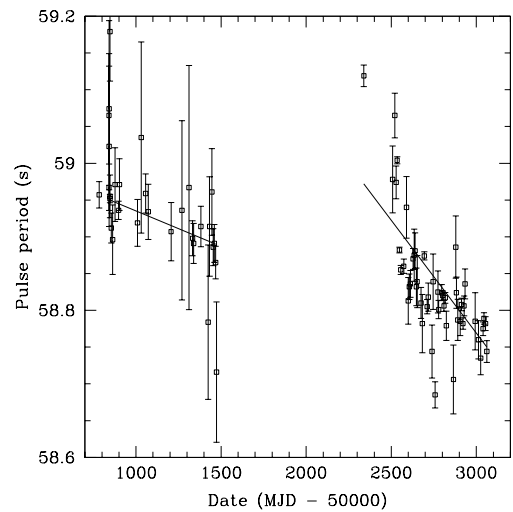}
\caption{An example of the spin period changes seen in one system, SXP59.0. The straight lines show the $\chi^2$ best fit to each data set.}
\label{fig1}
\end{center}
\end{figure}

The SMC has been the subject of extensive monitoring using the RXTE Proportional Counter Array (PCA) over the last 10 years. The PCA instrument has a Full Width Half Maximum (FWHM) field of view of $1\,^{\circ}$ and data in the energy range 3-10 keV were used. Most of the observations were pointed at the Bar region of the SMC where the majority of the known X-ray pulsar systems are located. Sources were identified from their pulse periods using Lomb-Scargle (Lomb 1976, Scargle 1982) power spectral analysis of the data sets. Laycock et al. (2005) and Galache et al. (2008) present full details of the data analysis approach that has been used to determine which pulsars were active during each observation. In their work, for each X-ray outburst, the pulse amplitude and history of the pulse periods were determined. Those results are used in this work.
Since a database of $\ge$10 years of observations of the SMC exists it was therefore possible to use these data to search for evidence of spin period changes in the systems.  The PCA is a collimated instrument, therefore interpreting the strength and significance of the signal depends upon the position of the source within the field of view. In all the objects presented here the target was assumed to be located at the position of the known optical counterpart. Only observations that had a collimator response $\ge$25\% and a detection significance of $\ge$99\% were used in this work.

A total of 24 sources were chosen for this study. In each case three possible measurements of period changes were obtained:

\begin{itemize}

\item Individual active periods lasting typically 50-500 days were studied and used to determine the $\dot{P}$ for a particular source data subset (referred to in this work as the Short$\dot{P}$). A simple best fit straight line to the pulse history plot was determined using a $\chi^2$ minimising technique. No attempt was made to fit more complicated profiles to the data, though in some cases higher order changes are suggested. An excellent example of the spin period changes seen in these systems is presented in Figure~\ref{fig1} which shows two outbursts from SXP59.0. Clearly both outbursts indicate an initially higher $\dot{P}$ which levels off as the activity period progresses, but only the weighted average is used in this work. One other factor that could also modify the spin period history would be Doppler-related changes. However, attempts to fit period histories with binary models have always proved difficult (but see, for example, Schurch et al, 2008, for one possible success) suggesting that the changes are dominated by accretion driven variability.

\item In addition to the Short$\dot{P}$, where possible a longer term value was determined for each source from the whole data set covering $\sim$10 years of observing  - referred to here as the Long$\dot{P}$. This typically included several periods of source activity with significant inactive gaps in between.

\item For many sources the orbital period is clearly apparent in the sequence of X-ray outbursts. For others, optical data from the Optical Gravitational Lensing Experiment (OGLE) project (Udalski, Kubiak and Szymanski 1997) have been used with good success to discover the orbital modulation - see Section 3.4 below for further discussion on this point.

\end{itemize}

It was not always possible to determine both a Short$\dot{P}$ and Long$\dot{P}$ for every source in this work due to several possible reasons; one being the observational coverage and another the activity history of the system. Details of the recorded spin period changes are given in Table~\ref{rxte}. Full records of the behaviour of each source may be found in Galache et al. (2008).

\begin{figure}
\begin{center}
\includegraphics[width=70mm,angle=-0]{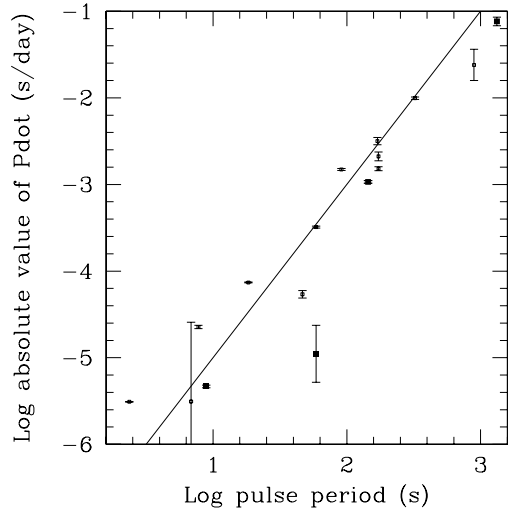}
\caption{The observed relationship between equilibrium spin period and the rate of spin change during an outburst (Short $\dot{P}$). The straight line shows a $\dot{P}$ = k$P^{2}$ relationship. The four points marked with a solid squares have positive $\dot{P}$ values (spin down), the rest are negative (spin up).}
\label{fig2}
\end{center}
\end{figure}

The strong link between the equilibrium spin period and the rate of change of spin period seen during outbursts is shown in Figure~\ref{fig2}. In this figure, the straight line represents $\dot{P}$ = k$P^{2}$ - as predicted for accretion from a disk on to a neutron star (see Equation 15 in Ghosh \& Lamb, 1979). It is interesting to note that the four spin-down systems (SXP8.80, SXP59.0, SXP144 \& SXP1323) fit comfortably on this relationship alongside the much larger number of spin-up systems.

To pursue the understanding of the accretion process further, a value for the X-ray luminosity, $L_x$, is needed during each outburst.
So the X-ray luminosity was calculated from the observed peak pulse amplitude in counts/pcu/s that occurred during the outburst that produced the Short$\dot{P}$ values listed in Table~\ref{rxte}. This amplitude is converted to luminosity assuming that 1 RXTE count/pcu/s = 0.4$\times 10^{37}$ erg/s at the distance of 60kpc to the SMC (though the depth of the sources within the SMC is unknown and \emph{may} affect this distance by up to $\pm$10kpc). The X-ray spectrum was assumed to be a power law with a photon index = 1.5 and an $N_{H}=6\times10^{20}cm^{-2}$. Furthermore it was assumed that there was an average pulse fraction of 33\% for all the systems and hence the correct total flux is 3 times the pulse component. Thus the values shown in Table~\ref{rxte} were determined using the relationship:
\\
\\
$L_X$ = 0.4 $\times\ 10^{37}$ $\times$ 3$R$  erg~s$^{-1}$
\\
\\ where $R$ = RXTE counts PCU$^{-1}$ s$^{-1}$

Note that though the values for $L_x$ obviously scale linearly with pulsed fraction, the neutron star mass determinations discussed below in Section 3.3 are very insensitive to any value between 10\% and 60\% for the pulsed fraction.

\begin{table*}
\begin{center}
\caption{Pulse period determinations. The X-ray luminosity (3--10 keV)comes from the peak amplitude of the pulsed modulation signal during an outburst - see text for details.The pulse period given in column 2 is the average value during the periods of activity used for measuring Short $\dot{P}$. Two systems (SXP59.0 and SXP172) have had two separate outbursts measured and so appear twice in the table.}
\label{rxte}
\begin{tabular}{ccccccccc}

\hline
Object  &Pulse&Error on&  Short $\dot{P}$&Error on& Long $\dot{P}$ &Error on& Peak $L_x$& Error on\\
name&period(s)&period(s)&(s/day)&Short $\dot{P}$&(s/day)&Long $\dot{P}$&($10^{37}$erg/s)&Peak $L_x$ \\
\hline

SXP2.37 & 2.372&0.001& -3.10$\times 10^{-6}$& 1.73$\times 10^{-8}$& -&- & 21.1&0.24 \\
SXP4.78 & 4.78&0.001&-&-& -2.50$\times 10^{-6}$ &3.97$\times 10^{-8}$& -&- \\
SXP6.85 & 6.851&0.002& -3.13$\times 10^{-6}$ &2.25$\times 10^{-5}$ &-& -& 1.44 &0.20\\
SXP7.78 & 7.785&0.002& -2.23$\times 10^{-5}$ &1.04$\times 10^{-6}$& +7.02$\times 10^{-6}$ &1.19$\times 10^{-7}$& 0.66 &0.12\\
SXP8.80 & 8.90&0.002 & +3.49$\times 10^{-6}$ & 1.59$\times 10^{-7}$&+4.57$\times 10^{-6}$&1.06$\times 10^{-7}$& 1.56 &0.11\\
SXP18.3 & 18.38&0.01& -7.40$\times 10^{-5}$ & 9.47$\times 10^{-7}$&-3.00$\times 10^{-5}$& 5.00$\times 10^{-7}$&1.66 &0.13\\
SXP46.6 & 46.5&0.05& -5.40$\times 10^{-5}$ & -5.76$\times 10^{-5}$& -6.94$\times 10^{-5}$& 1.14$\times 10^{-6}$& 0.60 &0.12\\
SXP51.0 & 51.1&0.04&  - &-& +3.39$\times 10^{-5}$ &4.83$\times 10^{-6}$& - &-\\
SXP59.0 & 58.9&0.02&+1.11$\times 10^{-5}$ &1.26$\times 10^{-5}$& -6.21$\times 10^{-5}$ &1.52$\times 10^{-6}$& 2.28& 0.24\\
SXP59.0 & 58.8&0.02& -3.22$\times 10^{-4}$ &8.70$\times 10^{-6}$& -6.21$\times 10^{-5}$ &1.52$\times 10^{-6}$& 2.16& 0.22\\
SXP82.4 & 82.4&0.03&  - &-& +5.23$\times 10^{-4}$ &7.21$\times 10^{-6}$& - &\\
SXP91.1 & 90.1&0.05& -1.49$\times 10^{-3}$ &4.11$\times 10^{-5}$& -5.48$\times 10^{-4}$ &5.68$\times 10^{-6}$& 1.68& 0.11\\
SXP144 & 144.5&0.1& +1.07$\times 10^{-3}$ &3.94$\times 10^{-5}$& +9.02$\times 10^{-4}$&2.44$\times 10^{-5}$& 0.72& 0.18\\
SXP152 & 151.5&0.2 &  -  &-& -4.47$\times 10^{-4}$ &8.34$\times 10^{-5}$& - &-\\
SXP169 & 167&0.5& -3.17$\times 10^{-3}$& 3.17$\times 10^{-4}$& -7.74$\times 10^{-4}$ &1.79$\times 10^{-5}$& 1.20& 0.22\\
SXP172 & 172.2&0.2& -1.52$\times 10^{-3}$ &8.54$\times 10^{-5}$& -4.24$\times 10^{-4}$ &1.71$\times 10^{-5}$& 0.36 &0.14\\
SXP172 & 171.6&0.2&$-2.11\times 10^{-3}$ &2.62$\times 10^{-4}$& -4.24$\times 10^{-4}$  &1.71$\times 10^{-5}$& 0.60 &0.12\\
SXP202A& 202.5&0.3&  -  &-& -2.69$\times 10^{-4}$ &3.33$\times 10^{-5}$& - &-\\
SXP293 & 294&0.3 & - &-& -3.32$\times 10^{-4}$ &3.00$\times 10^{-4}$& - &-\\
SXP323 & 319&0.5& -9.90-$\times 10^{-3}$ &3.39$\times 10^{-4}$& -2.72$\times 10^{-3}$  &5.70$\times 10^{-5}$& 0.84 &0.14\\
SXP342 & 342&1.0&-&  - & +2.46$\times 10^{-3}$ &1.92$\times 10^{-4}$& - &-\\
SXP504 & 503&1.5&-& - & +8.08$\times 10^{-4}$ &1.91$\times 10^{-4}$& - &-\\
SXP701 & 696&2.5&-& - & +9.45$\times 10^{-4}$ &1.02$\times 10^{-3}$& - &-\\
SXP756 & 755 &3.5&-& - & -2.01$\times 10^{-3}$ &7.67$\times 10^{-4}$& - &-\\
SXP892 & 895&5.0& -2.40$\times 10^{-2}$ &1.23$\times 10^{-2}$& -1.97$\times 10^{-3}$ &1.82$\times 10^{-3}$& 0.48&0.12\\
SXP1323 & 1325&8.0& +7.63$\times 10^{-2}$ &9.23$\times 10^{-3}$& - & - & 2.88 &0.30\\

\hline
\end{tabular}
\end{center}
\end{table*}

\begin{figure}
\begin{center}
\includegraphics[width=90mm,angle=-0]{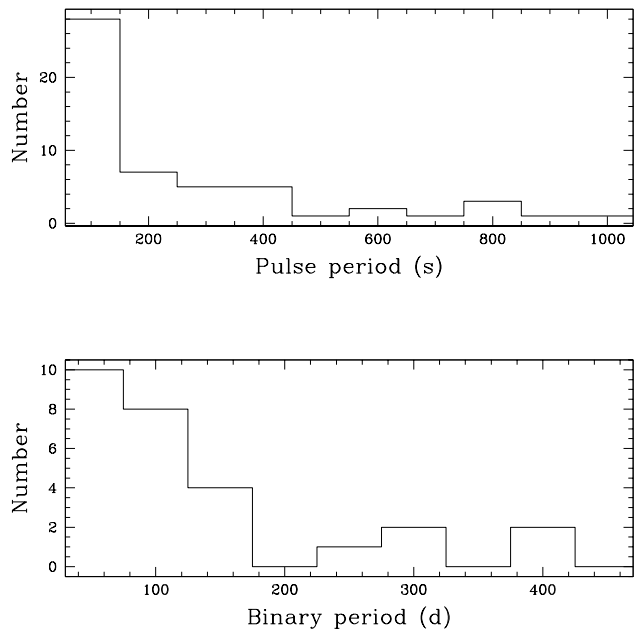}
\caption{The histograms of the observed pulse periods (upper) and binary periods (lower) for SMC systems. The majority of these systems are described in Galache et al. (2008), but others from the published literature have also been included.}
\label{fig3}
\end{center}
\end{figure}

Figure~\ref{fig3} (upper panel) shows a histogram of all the known pulse periods for accreting pulsar systems in the SMC (see Galache et al, 2008 for details of the vast majority of the sources). Because of the length of the each observation (typically 10ks - 15ks) there is undoubtedly an instrumental cut-off starting around $\ge$1ks affecting the ability to detect longer periods. The sharpness of this cutoff is partially a function of the pulse profile. At the short period end, the data were regularly searched down to periods of 0.5s (Galache et al, 2008) and should be complete down to that number. The lower panel in Figure~\ref{fig3} shows the distribution of known binary periods determined from either the sequence of X-ray outbursts and/or a coherent modulation of the optical light from the system.

\begin{figure}
\begin{center}
\includegraphics[width=70mm,angle=-0]{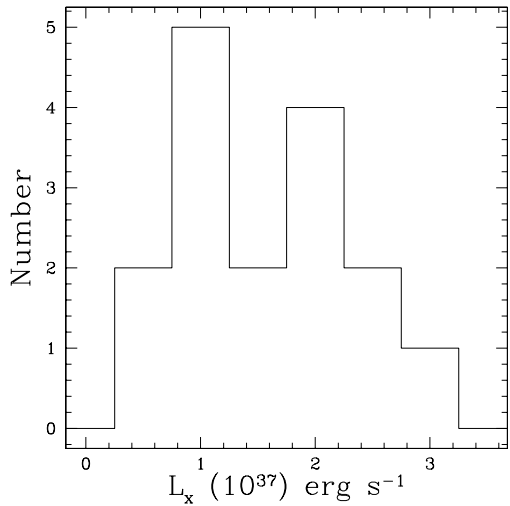}
\caption{The histogram of the estimated peak X-ray luminosities for the outbursts discussed in this work. The exceptionally large outburst from SXP2.37 is excluded from this plot.}
\label{fig4}
\end{center}
\end{figure}

Figure~\ref{fig4} shows a histogram of the peak outbursts observed from the systems studied in this work. The significantly larger outburst seen from SXP2.37 is excluded from this plot - this one outburst may well be tending towards a Type II outburst.

\section{Discussion}

\subsection{Reverse torques}
It is interesting to note that of the 15 systems for which short term $\dot{P}$ changes were measured, 4 (SXP8.80, SXP144, SXP59.0 and SXP1323) show spin-downs during outburst rather than spin-ups. Such reverse torques have been reported before and even transitions from spin-up to spin-down observed {for example, 4U1626-67 (Bildsten et al, 1997)}. No such transitions are observed in the sample presented here, but the numbers do give a clear indication of the frequency with which these reverse torques occur. In addition, it is obviously a phenomenon that is not restricted to a particular spin period regime since our three object have periods ranging from 8.9s to 1325s. Five further systems show long term spin-down changes - see  Table~\ref{rxte}. However, all these other systems show spin-up during outbursts and the long-term change may be attributed to the gradually slowing down of the neutron star in the absence of repeated accretion episodes, rather than reverse torques.

\begin{figure}.
\begin{center}
\includegraphics[width=80mm,angle=-0]{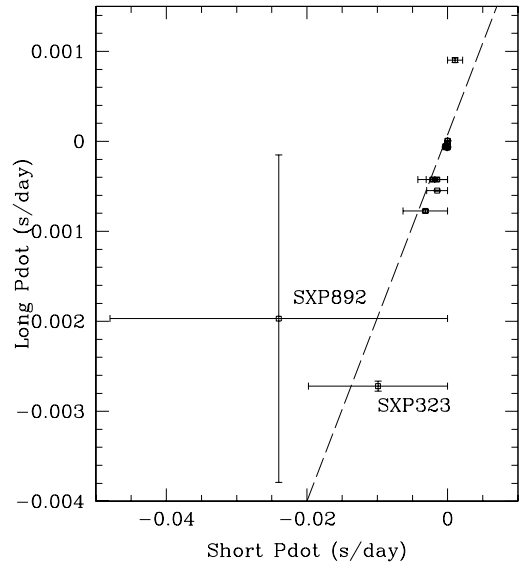}
\caption{The relationship between the absolute values of the short ($\sim$couple of months) spin changes and the longer term changes ($\sim$10 years). All 11 sources from Table 1 with both short \& long term $\dot{P}$ values are included. The dashed line shows the simple relationship Long$\dot{P}$ = 0.2 $\times$ Short$\dot{P}$.}
\label{fig5}
\end{center}
\end{figure}

\subsection{Spin period evolution}
The relationship between short spin period changes (seen during an outburst of typically 1-2 months) and the long term period changes seen over $\sim$10 years worth of study is shown in Figure~\ref{fig5}. The sources broadly follow a relationship indicating that Long$\dot{P}$ = 0.2 $\times$ Short$\dot{P}$. So individual outbursts typically spin up (or down) a neutron star $\sim$5 times faster than the longer, time-averaged changes. In reality, the neutron star is subjected to a series of ``kicks'' during each short outburst, followed by an intervening recovery period. This effect is illustrated clearly in Figure~\ref{fig1}.

\subsection{Neutron star masses}

\begin{figure*}
\begin{center}
\includegraphics[width=140mm,angle=-0]{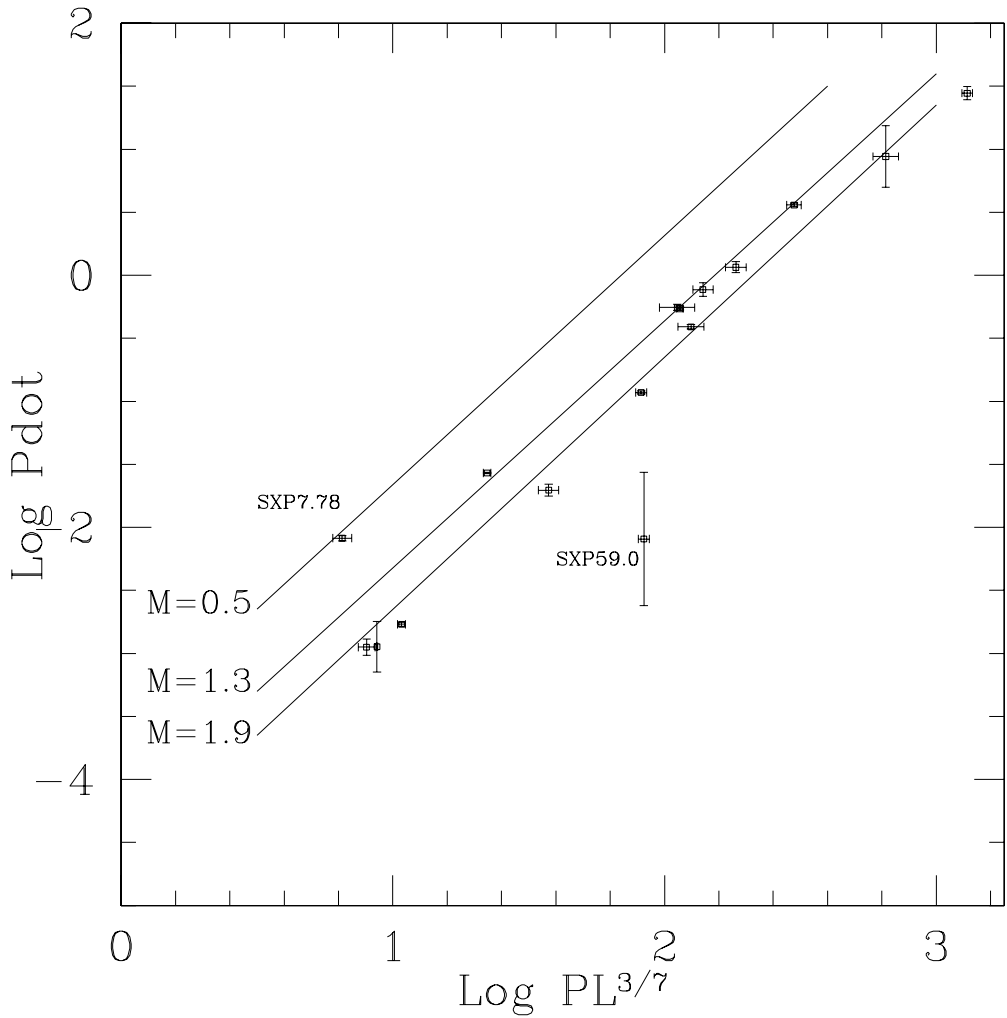}
\caption{Adapted from Figure 10 of Ghosh \& Lamb (1979) for disk accretion on to neutron stars with a magnetic moment of $\mu$=0.48. All 16 outbursts with a measurable Short$\dot{P}$ from Table 1 are included. The three parallel lines represent three different possible neutron star masses.}
\label{fig6}
\end{center}
\end{figure*}

Ghosh \& Lamb (1979) established the relationships for accretion onto neutron stars formulating the now well known relationship between $\dot{P}$ and X-ray luminosity (see Equation 15 in their paper). This relationship was re-iterated by Joss \& Rappaport (1984) and others, with all sets of authors comparing the theoretical predictions with available data from galactic accreting pulsar systems. The quality of the observational data used was very variable, with many objects represented simply by upper limits. In this work the consistently higher quality of the measurements allows us to compare theory and observation much more precisely. This comparison is shown in Figure~\ref{fig6}, where it is immediately obvious that the data provide strong support for the models over two orders of magnitude. 
The probability of getting this level of correlation with an uncorrelated data set is 2.5$\times 10^{-9}$.

From the results presented here there is a strong support for the average mass of neutron stars in these systems to lie between 1.3\mbox{M$_\odot$} and 1.9\mbox{M$_\odot$}. In fact, if the sample shown in Figure~\ref{fig6} is used, then the average mass of the neutron stars in the SMC sample is found to be 1.62$\pm$0.29M$_\odot$. The only significant deviation from the average value is that of SXP7.78 (=SMC X-3) which strongly suggests a much lower neutron star mass for that system. SXP59.0 has the largest error bars of all the points and lies $\sim$2$\sigma$ away from the M=1.9M$_\odot$; hence is probably not significant measurement of a large neutron star mass.

\subsection{Implications for HMXB evolution}

There are currently 56 systems with known pulse periods, but only 27 with confirmed binary periods. Many of the binary periods have been determined or confirmed from OGLE III long term data which frequently show evidence of modulation believed to be at the binary period. Results such as those for SXP46.6 (McGowan et al, 2008) demonstrate the strength of this link by revealing regular optical outbursts coincident with X-ray outbursts. However, for many of the SMC systems the binary period has yet to be confirmed and, in particular, the longer term periods ($\ge$1-2 years) are increasingly difficult to ascertain. This may be due to either missing X-ray outbursts or/and a lack of optical modulation due, perhaps, to the size and shape of the orbit.

However the period histogram should reflect accurately the distribution of pulse periods found in accreting pulsar systems (at least in the SMC). There is obviously a strong preference to forming systems with spin periods of $\le$100s. Since the Corbet Diagram (Corbet et al., 2008) provides strong evidence that the equilibrium spin period is driven by the orbital period, then this spin period result probably reflects the relative difficulty in long orbital period systems surviving the supernova explosion that produces the neutron star. This is supported by the binary period distribution seen in the lower panel of Figure~\ref{fig3} (but with all the observational caveats mentioned above). It is also worth noting that there is no evidence from the Corbet diagram that SMC sources show any difference from their counterparts in the Milky Way (Corbet et al., 2008).

\section{Conclusions}

This work presents results from $\sim$10 years of X-ray monitoring of the pulse period histories of 24 HMXB systems in the SMC. This homogenous group of objects provides an excellent test bed for accretion theory and, in general, the results are very supportive of current models. Since current models have all been developed based upon Milky Way systems, this work strongly supports the conclusion that there is little, or no difference between HMXB behaviour in the SMC and the Galaxy.

\section{Acknowledgements}

We are grateful to Lee Townsend for help with some of the data analysis. In addition, we are grateful to the helpful comments of the referee which enhanced the robustness of this paper.

\bsp

\label{lastpage}

\end{document}